\documentclass[a4,12pt]{article}

\usepackage[latin1]{inputenc}

\newcommand{\jmu}{\sigma}

\newcommand{\beijing}[1]{}

\newcommand{\tU}{\tilde U}

\newcommand{\MUone}{M/\mathrm{U(1)}}%
\newcommand{\Uone}{\mathrm{U(1)}}%

\usepackage{color}
\usepackage{graphicx}  

\usepackage{psfrag}

\usepackage{tocbibind} 

\usepackage{epsf}
\usepackage{epic}
\usepackage{eepic}

\usepackage{epsfig}
\usepackage{wrapfig}
\usepackage{amsmath,amssymb,amsfonts}
\usepackage[mathscr]{eucal}
\usepackage{citesort}
\usepackage[active]{srcltx}
\usepackage{url}
\usepackage{fancybox}
\usepackage{ntheorem}

\usepackage{mhequ}

\newcommand {\twist} {\lambda}

\newcommand{\myomega}{v}

\newcommand{\beq}{\begin{equation}}

\newcommand{\FS}       
                  {F}

\newcommand{\HS} 
       {H_{\mbox{\scriptsize volume}}}

{\ptc{this should be removed in the oberwolfach version}}%

\newcommand{\mcA}{{\mycal A}}%
\newcommand{\eeal}[1]{\label{#1}\end{eqnarray}}
\newcommand{\bed}{\begin{deqarr}}
\newcommand{\eed}{\end{deqarr}}
\newcommand{\bedl}[1]{\begin{deqarr}\label{#1}}
\newcommand{\eedl}[2]{\arrlabel{#1}\label{#2}\end{deqarr}}

\newcommand{\bel}[1]{\begin{equation}\label{#1}}
\newcommand{\bea}{\begin{eqnarray}}
\newcommand{\bean}{\begin{eqnarray}\nonumber}
\newcommand{\beal}[1]{\begin{eqnarray}\label{#1}}
\newcommand{\eea}{\end{eqnarray}}

\newcommand{\qed}{\hfill $\Box$ \medskip}
\newcommand{\proof}{\noindent {\sc Proof:\ }}
\newcommand{\be}{\begin{equation}}
\newcommand{\eeq}{\end{equation}}
\newcommand{\ee}{\end{equation}}
\newcommand{\beqa}{\begin{eqnarray}}
\newcommand{\eeqa}{\end{eqnarray}}
\newcommand{\beqan}{\begin{eqnarray*}}
\newcommand{\eeqan}{\end{eqnarray*}}
\newcommand{\ba}{\begin{array}}
\newcommand{\ea}{\end{array}}

\newcommand{\mcW}{{\mycal W}}

\newtheorem{Theorem} {\sc  Theorem\rm} [section]

\newtheorem{Lemma} [Theorem] {\sc  Lemma\rm}

\theorembodyfont{\upshape}
\newtheorem{Remark}[Theorem]{\sc Remark\rm}
\newtheorem{remark}[Theorem]{\sc Remark\rm}

\theoremsymbol{\ensuremath{\diamondsuit}}
\newcommand{\fcoco}{\small}
\theorembodyfont{\fcoco} \theoremseparator{} \theoremindent0.5cm

\theoremstyle{nonumberplain} \theorembodyfont{\fcoco}
\theoremseparator{} \theoremindent0.5cm

\DeclareFontFamily{OT1}{rsfs}{}
\DeclareFontShape{OT1}{rsfs}{m}{n}{ <-7> rsfs5 <7-10> rsfs7 <10->
rsfs10}{} \DeclareMathAlphabet{\mycal}{OT1}{rsfs}{m}{n}

{\catcode `\@=11 \global\let\AddToReset=\@addtoreset}
\AddToReset{equation}{section}
\renewcommand{\theequation}{\thesection.\arabic{equation}}

\newcounter{mnotecount}[section]

\renewcommand{\themnotecount}{\thesection.\arabic{mnotecount}}

\newcommand{\mnote}[1]
{\protect{\stepcounter{mnotecount}}$^{\mbox{\footnotesize
$
\bullet$\themnotecount}}$ \marginpar{
\raggedright\tiny\em
$\!\!\!\!\!\!\,\bullet$\themnotecount: #1} }

\newcommand{\warn}[1]
{\protect{\stepcounter{mnotecount}}$^{\mbox{\footnotesize
$
\bullet$\themnotecount}}$ \marginpar{
\raggedright\tiny\em $\!\!\!\!\!\!\,\bullet$\themnotecount: {\bf
Warning:} #1} }

\newcommand{\R}{\mathbb R}

\newcommand{\eq}[1]{(\ref{#1})}

\newcommand{\Mext}{M_\ext}
\newcommand{\ext}{\mathrm{ext}}

\newcommand{\ptc}[1]{\mnote{{\bf ptc:}#1}}

\newcommand{\mcL}{{\mycal L}}

\newcommand{\beqar}{\begin{deqarr}}
\newcommand{\eeqar}{\end{deqarr}}

\newcommand{\beaa}{\begin{eqnarray*}}
\newcommand{\eeaa}{\end{eqnarray*}}

\newcommand{\tr}{\mbox{tr}}

\title{A Dain Inequality with charge}

\author{Jo\~ao Lopes Costa
\\
Lisbon University Institute -- ISCTE
\\
Mathematical Institute and Magdalen College, Oxford
 }

\begin{document}
\maketitle

\begin{abstract}
We prove an upper bound for angular-momentum and charge in terms of
the mass for electro-vacuum asymptotically flat axisymmetric initial
data sets with simply connected orbit space.
\end{abstract}

%
\section{Introduction}
\label{Sintro}

Gravitational collapse involving {\em suitable} matter is
expected~\cite{Penrose:1969,Wald:1997,Dain:2005} to {\em
generically} result in the formation of an event horizon whose
exterior solution approaches a Kerr-Newman metric asymptotically
with time, here we are assuming that the exterior region becomes
electro-vacuum. Then, the characteristic inequality
\bel{21VI.1}
 m\ge \sqrt{\frac{|\vec J|^2}{m^2}+Q_E^2+Q_B^2} \;
 \ee
relating the ADM mass, ADM angular momentum and charges of such
black-holes should also be valid asymptotically with time. Imposing
electro-vacuum at large distances, the mass and Maxwell charges are
conserved quantities and the same is true for angular momentum if
one further assumes axisymmetry. Consequently,
inequality~\eqref{21VI.1} should hold for axisymmetric initial data
for such a collapse. Besides their own intrinsic interest, results
establishing such inequalities provide evidences in favor of this
``current standard picture of gravitational
collapse"~\cite{Dain:2005}, which is based upon {\em weak cosmic
censorship} and a version of {\em black hole uniqueness}
considerably stronger than the ones
available~\cite{ChrusCosta:2008,Costa:2009}.

 Dain~\cite{Dain:2005,Dain:2006}, besides providing the previous Penrose-like heuristic argument, proved an
upper bound for angular-momentum in terms of the mass for a
class of maximal, vacuum, axisymmetric initial data
sets.\footnote{The analysis of~\cite{Dain:2006} has been
extended in~\cite{CLW} to include all vacuum axisymmetric initial
data, with simply connected orbit space, and manifolds which
are asymptotically flat in the standard sense, allowing
moreover several asymptotic ends.}
Recently a generalized Dain's inequality including electric and
magnetic charges was obtained in~\cite{ChrusCosta:2009}; there the
proof of the main result, based on the methods of~\cite{CLW}, was
only sketched. The aim of this work is to provide a complete proof
of this charged Dain inequality while simplifying the methods
of~\cite{CLW}.

 More precisely, consider
a three dimensional electro-vacuum smooth initial data set $(M,g,K,
E, B)$,  where $M$ is the union of a compact set and of two
asymptotically flat regions  $M_1$ and $M_2$. Here $g$ is a
Riemannian metric on $M$, $K$ is the extrinsic curvature tensor, $E$
is the electric field and $B$ the magnetic one; electro-vacuum means
that the constraints~\eqref{EMcost} are satisfied with $E$ and $B$
both divergence-free.  We suppose that the initial data set is
\emph{axisymmetric}, by which we mean that it is invariant under an
action of $\Uone$, and maximal: $\tr_g K =0$. It is further assumed
that $\MUone$ is simply connected, so that the results
of~\cite{ChUone} can be used. The notion of asymptotic flatness is
made precise in \eq{falloff1} and \eq{KFfalloff1}, where moreover
$k\ge 6$ needs to be assumed when invoking~\cite{ChUone}.  We will
prove the following result:

\begin{Theorem}
 \label{TM1}
Under the conditions just described, let $m$, $\vec J$, $Q_E$ and
$Q_B$ denote respectively the ADM mass~\eqref{mf}, the ADM angular
momentum~\eqref{Jvanish} and the total electric and magnetic
charges~\eqref{charges} of $M_1$. Then
\bel{21VI.2}
 m\ge \sqrt{\frac{|\vec J|^2}{m^2}+Q_E^2+Q_B^2} \;.
\ee
\end{Theorem}

\begin{Remark}
We expect the equality to be attained only for the magnetically and
electrically charged extreme Kerr-Newman space-times, which do not
satisfy the hypotheses of Theorem~\ref{TM1}. Indeed, any spacelike
manifold in an extreme Kerr-Newman space-time is either incomplete,
or contains a boundary, or a singularity, or an asymptotically
cylindrical end.
\end{Remark}

\begin{Remark}
\label{J=0} If $M$ contains only one asymptotic flat end and
$\partial M=\emptyset$ we have $Q_E=Q_B=0$ and, under the
supplementary hypothesis~\eqref{29VI.1}, we also have $\vec J=0$
(see \eq{Jvanish} below). Whence our interest in initial data sets
containing two ends. In fact one expects our main result to
generalize to several ends along the lines of~\cite{CLW}, but a
proof of this lies beyond the scope of this work.
\end{Remark}

\begin{Remark}
The proof applies to Einstein--Abelian Yang-Mills fields
configurations, giving in this case
\bel{21VI.2}
 m \ge \sqrt{\frac{|\vec J|^2}{m^2}+(\sum_i Q_{E_i})^2+(\sum_i Q_{B_i})^2}
 \;,
\ee
where the $Q_{E_i}$'s and the $Q_{B_i}$'s are the electric and
magnetic charges associated with the $i$'th Maxwell field.
\end{Remark}

A slightly more general version of Theorem~\ref{TM1} can be found in
Theorem~\ref{TM2} below. The reader should also note an inequality
relating area, angular momentum, and charge, proved for {\em
stationary} Einstein-Maxwell black holes
in~\cite{HennigAnsorgCederbaumEM}, as well as  the discussion of the
Penrose inequality in electrovacuum of~\cite{WeinsteinYamada}.

\section{Mass, angular momentum and charge inequalities}
 \label{SAmci}

Recall that an \emph{asymptotically  flat end} is a region
$\Mext\subset M$ diffeomorphic to $\R^3\setminus B(R)$, where $B(R)$
is a coordinate ball of radius R, such that in coordinates on
$\Mext$ obtained from $\R^3\setminus B(R)$ we
have, for some $k\geq1$,%
\footnote{We write $f=o_k(r^{-\alpha})$ if the limits $
  \lim_{r\to\infty}r^{\alpha+\ell}\partial_{k_1}\ldots\partial_{k_\ell}
f$ vanish for all $ 0\le \ell \le k $, and $f=O_k(r^{-\alpha})$
if there exists a constant $C$ such that $
  |r^{\alpha+\ell}\partial_{k_1}\ldots\partial_{k_\ell}
f|\le C$  for all $ 0\le \ell \le k $.}
\bel{falloff1} g_{ij}=\delta_{ij}+o_k(r^{-1/2})
 \;, \ \partial_k g_{ij} \in L^2(\Mext)\;, \quad
 K_{ij}=O_{k-1}(r^{-\beta}) \;,\  \beta > \frac 52\;.
\ee
(The asymptotic conditions on $g$ arise from the requirement of well
defined ADM mass, with the integrability condition satisfied if,
e.g., $\partial_k g_{ij}=O(r^{-\alpha-1})$, for some
$\alpha>1/2$~\cite{ChUone}. The restriction on the decay rate of $K$
above arose already in the vacuum case, and can be traced back to
the unnumbered equation after (2.37) in~\cite{CLW}.)

The electric and magnetic fields $E$ and $B$ are
the orthogonal projections to $TM$ of their space-time
analogues
\bel{electromagnetic fields}
E^{\mu}=F^{\mu}{}_{\nu}n^{\nu}\;,\quad
B^{\mu}=*F^{\mu}{}_{\nu}n^{\nu}\;, \ee
where $F$ is the Maxwell two-form, and where $n$ is a unit
normal to $M$, when embedded in an electro-vacuum space-time.%
\footnote{The existence of an electro-vacuum and axisymmetric
evolution of the data follows from its smoothness
by~\cite{ChoquetBook} and~\cite{Chrusciel-completeKV}. This will, in
particular, allow us to use the (space-time) computations
in~\cite{Weinstein:1996}.}
 We assume that in
the manifestly asymptotically flat coordinates we have
\bel{KFfalloff1}
 E^i=O_{k-1}(r^{-\gamma-1}) \;, \quad
 B^i=O_{k-1}(r^{-\gamma-1})\;\;,\; \gamma>3/4
 \;.
\ee
%

The Einstein-Maxwell scalar constraint equation reads, for maximal
initial data,
\bel{EMcost} ^{(3)}R = 16 \pi \mu+ |K|_g^2 + 2\Big(|E|_g^2+|B|^2_g
\Big)
 \;,
\ee
where the function $\mu\ge 0$ represents the non-electromagnetic
energy density and $|\cdot|_g$ denotes the norm of a vector with
respect to the metric $g$

To obtain our inequality we start by bounding \eq{EMcost} from
below, as follows. By~\cite{ChUone} there exists a coordinate
system, with controlled asymptotic behaviour, in which the metric
takes the form
\begin{equation} \label{axmet2}
g = e^{-2U+2\alpha} \left(d\rho^2 + dz^2 \right) + \rho^2 e^{-2U}
\left(d\varphi + \rho\: W_{\rho} d\rho + W_z dz \right)^2 \, ;
\end{equation}
such coordinates are global in $M_1$, with $M_2$ being represented
by the ``puncture" $\{\rho=z=0\}$.

 Consider an orthonormal frame $e_i$ such that $e_3$ is proportional
to the rotational Killing vector field
$$
 \eta:=\partial_\varphi
 \;.
$$
Let $\theta^i$ denote the dual co-frame; for definiteness we
take
$$
\theta^1 =e^{-U+\alpha} d\rho\;, \quad \theta^2 = e^{-U+\alpha}
dz\;, \quad \theta^3 = \rho e^{-U} \left(d\varphi + \rho W_{\rho}
d\rho + W_z dz \right) \; .
$$
We assume that the initial data are invariant under the flow of
$\eta$; this implies the space-time equations $\mcL_\eta F=0$ and
$\mcL_\eta * F=0$,  where a star stands for the Hodge dual, and
$\mcL$ denotes a Lie-derivative. Then, Maxwell's equations
\bel{Maxeq}
dF=d*F=0\;,
\ee
together with the hypothesis of simple-connectedness of $\MUone$,
imply the existence of functions $\chi$ and $\psi$ such that
\bel{Ernst}
 \partial_\alpha \chi=F_{\mu\alpha} \eta^\mu \;,\qquad
 \partial_\alpha \psi=*F_{\mu\alpha} \eta^\mu  \;.
\ee
In the orthonormal basis $\{n, e_i\}$ we have
$$
F_{\mu\nu}=
\left(\begin{array}{cccc}
\;\;\;0      &  -E_1       &  - E_2 &    -E_3 \\
\;\;\;E_1    & \;\; 0            & \;\;\; B_3 & -B_2 \\
\;\;\;E_2    &  - B_3 & \;\; 0   & \;\;\; B_1 \\
\;\;\;E_3    &  \;\;\;B_2        &  - B_1     & \;\;0
\end{array}\right)
\;,
$$
therefore
$$(\partial_\alpha \chi)=\sqrt{g_{\varphi\varphi}}\left(F_{3\alpha}\right)=\rho e^{-U}(E_3,B_2,-B_1,0)\;,$$
which together with the analogous expression for $(\partial_\alpha
\psi)$ yields
\bean
   |E|_g^2+|B|^2_g
 & = &
  \frac {e^{2U}}{\rho^2}\Big((\partial_n \chi)^2 +|D\chi|_g^2 +
  (\partial_n \psi)^2 +|D\psi|_g^2 \Big)
  \\
  & \ge &
  \frac {e^{2U}}{\rho^2}\Big(|D\chi|_g^2 +
   |D\psi|_g^2 \Big)
  \nonumber
  \\
  & = &
  \frac {e^{4U-2\alpha}}{\rho^2}\Big((\partial_\rho\chi)^2 +(\partial_z\chi)^2 +
   (\partial_\rho\psi)^2+
   (\partial_z\psi)^2 \Big)
  \;,%
\eeal{EBlb}
where $\partial_n$ denotes the derivative in the direction normal to
the initial data hypersurface.
\begin{remark}
\label{stationary}
 For a stationary development of the data, if we
let $\tau$ denote the stationary vector, the field equations imply
the integrability conditions~\cite{Weinstein:1996,Costa:2009}
\bel{intcond} d\tau^{\flat}\wedge\tau^{\flat}\wedge
\eta^{\flat}=d\eta^{\flat}\wedge\tau^{\flat}\wedge \eta^{\flat}=0\;.
\ee
It then follows that $n$ is spanned by the Killing vectors $\tau$
and $\eta$, and~\eqref{EBlb} becomes an equality.
\end{remark}
 Writing
$$
4\pi T_{\mu\nu}\eta^{\mu}
=F_{\mu\alpha}F_{\nu}{}^{\alpha}\eta^{\mu}-\frac{1}{4}F_{\alpha\beta}F^{\alpha\beta}\eta_\nu
= \partial_{\alpha}\psi F_{\nu}{}^{\alpha}-\frac{1}{4}
F_{\alpha\beta}F^{\alpha\beta}\eta_\nu
 \;,
$$
we are now able to justify Remark~\ref{J=0}. First, the vanishing of
the electric and magnetic charges is an immediate consequence of
their definition
\bel{charges} Q_E=-\frac{1}{4\pi}\int_{S_{\infty}}*F\;,
 \quad
 Q_B=\frac{1}{4\pi}\int_{S_{\infty}}F\;,
\ee
and Maxwell's equations~\eqref{Maxeq}. Next, the vanishing of
angular momentum will be established under the supplementary
condition that
\bel{29VI.1}
 \psi F_{\mu\nu} = o(r^{-2})
 \;.
\ee
The symmetry of the problem implies that $\vec J$ is aligned along
the axis of rotation. Letting $J_z$ denote the component of
angular-momentum along the rotation axis, by the Komar identity we
obtain, recall that here we are assuming that $\partial
M=\emptyset$,
\begin{eqnarray}
 \nonumber
 16\pi J_z & = &
 \int_{S_\infty}\nabla^{\mu}\eta^{\nu}dS_{\mu\nu}=\frac{1}{2}\int_M \nabla_{\mu}\nabla^{\mu}\eta^{\nu}dS_{\nu}
\\
 \nonumber
& = &\frac{1}{2} \int_M R_{\mu}{}^{\nu}\eta^{\mu}dS_{\nu}=4\pi\int_M T_{\mu}{}^{\nu}\eta^{\mu}dS_{\nu}
\\
& = &
 \nonumber
-\int_{M}
\psi\underbrace{\nabla_{\alpha}F^{\nu\alpha}}_0dS_{\nu}+\int_M
\nabla_{\alpha}(\psi F^{\nu\alpha})dS_{\nu}-\frac{1}{4}\int_{M}
F_{\alpha\beta}F^{\alpha\beta}\eta^\nu dS_{\nu}
\\
& = &   2\int_{S_{\infty}} \psi F^{\nu\alpha}dS_{\nu\alpha}
-\frac{1}{4}\int_{M}F_{\alpha\beta}F^{\alpha\beta}\underbrace{\eta^{\mu}n_{\mu}}_0d^3\mu_g=0
 \;,
 \label{Jvanish}
\end{eqnarray}
where we have used the fact that $\eta$ is tangent to $M$, and
where the first integral in the last line vanishes by
\eq{29VI.1}.

As discussed in~\cite{Dain:2006}, in vacuum the one-form%
\footnote{We take this opportunity to point out a factor of 2
missing in the left-hand-side of Eq.~(2.6) in~\cite{CLW}, which
affects numerical factors in some subsequent equations, but has
no other consequences.}
 \bean
 \label{lambda}
 \twist  &:=& 2
 \epsilon_{ijk}K^j{}_\ell \eta^k \eta^\ell dx^i
 \\
 \nonumber
 &=& 2
  \epsilon({\partial_A,\partial_B,\partial_\varphi})K(dx^B,\partial _\varphi) dx^A
 \\
 &=&
 2  g(\eta,\eta)\epsilon({e_a,e_b,e_3})K(\theta^b,e_3) \theta^a
\eeal{twist}
is closed. Here, as before, the upper case indices $A,B=1,2$
correspond to the coordinates $(\rho, z)$, while the lower case
indices $a,b=1,2$ are frame indices. In electro-vacuum we have
instead (see, e.g.,~\cite{Weinstein:1996})
\bel{twist2}
  d\Big(\twist-2(\chi d\psi - \psi d\chi)\Big)=0
 \;,
\ee
%
and, since we have assumed that $\MUone$ is simply connected, there
exists a function $\myomega$ such that
\bel{twistex}
  \twist  = 2 (d\myomega + \chi d\psi - \psi d\chi)
 \;.
\ee
Then, writing $K_{b3}$ for $K(e_b,e_3)$, and using
$\epsilon_{ab}:=\epsilon(e_a,e_b,e_3)\in \{0,\pm 1\}$, we have
$$
2\rho^2 e^{-2U}(K_{23}\theta^1- K_{13}\theta^2)= \twist_\rho  d\rho +
\twist_z  dz
 \;;
$$
equivalently
\bel{Komeq} K_{13}=-\frac{e^{3U-\alpha}}{2\rho^2}\twist_z \;,
 \quad
 K_{23}=\frac{e^{3U-\alpha}}{2\rho^2}\twist_\rho \;,
\ee
so that
%
\bel{Klb}
e^{2(\alpha-U)}\,|K|_g^2 \ge 2
 e^{2(\alpha-U)}
(K_{13}^2+K_{23}^2) = \frac {e^{4U}} {2\rho^4} |\twist
|^2_\delta
 \;.
\ee

In~\cite{ChUone}
(compare~\cite{Brill59,GibbonsHolzegel,Dain:2006}) it has been
shown that
\bean m&=& \frac{1}{16 \pi} \int \Big[\phantom{}^{(3)}R +
\frac{1}{2} \rho^2
  e^{-4\alpha+2U}\left(\rho W_{\rho, z} - W_{z,\rho}\right)^2 \Big] e^{2(\alpha -U)} d^3 x
 \\
 & & +
\frac{1}{8\pi}\int \left(D U\right)^2 d^3x
 \label{mf}
\\
& \ge & \frac{1}{16 \pi} \int \Big[\phantom{}^{\, (3)}R
e^{2(\alpha-U)}+   2\left(D U\right)^2 \Big]d^3x \, . \label{mf2}
\eea
Inserting \eq{EBlb} and \eq{Klb} into \eq{mf2} we obtain
\bel{Mainineq} m \ge \frac{1}{8 \pi} \int \Big[ \left(D U\right)^2
+\frac {e^{4U}} {\rho^4} \left(D \myomega+\chi D\psi - \psi D \chi
\right)^2 +\frac {e^{2U}} {\rho^2} \left((D  \chi)^2+
(D\psi)^2\right)\Big]d^3x
 \, ,
\ee
where, from now on, we use the symbol $Df$ to denote the gradient of
a function $f$ with respect to the flat metric
$\delta=d\rho^2+dz^2+\rho^2 d\varphi^2$, and we will use both
$(v)^2$ and $|v|^2$ in alternative to $|v|_{\delta}^2$, the square
of the norm of a vector $v=v^A\partial_A$ with respect to $\delta$.
\begin{remark}
\label{m=I} As a consequence of~\cite{Dain:2005} and
Remark~\ref{stationary} we see that for {\em stationary} data with
vanishing non-electromagnetic energy density ($\mu=0$)
$$
m = \frac{1}{8 \pi} \int \Big[ \left(D U\right)^2 +\frac {e^{4U}}
{\rho^4} \left(D \myomega+\chi D\psi - \psi D \chi \right)^2 +\frac
{e^{2U}} {\rho^2} \left((D  \chi)^2+ (D\psi)^2\right)\Big]d^3x\;.
$$
\end{remark}

It follows from \eq{Ernst} that $\psi$ and $\chi$ are constant
on each connected component $\mcA_j$ of the ``axis"
$$
 \mcA:=\{\rho=0\}\setminus\{z=0\}\;;
$$
\eq{twistex}-\eq{Komeq} then
show that so is $\myomega$. We set
\bel{AxisValue}
 \myomega_j:= \myomega|_{\mcA_j}
 \;, \quad \psi_j:=\psi|_{\mcA_j}
 \;, \quad \chi_j:=\chi|_{\mcA_j}
 \;, \quad j=1,2
 \;,
\ee
where $\mcA_1=\{\rho=0\}\cap\{z<0\}$ and
$\mcA_2=\{\rho=0\}\cap\{z>0\}$.

 We have the following, from which Theorem~\ref{TM1}
immediately follows:

\begin{Theorem}
 \label{TM2}
Let $(M,g,K,   v,   \chi,\psi)$ be a three dimensional smooth data
set invariant under an action of $\Uone$, where $M$ is the union of
a compact set and of two asymptotically flat  regions $M_1$ and
$M_2$, in the sense of \eq{falloff1}, and where $v,\psi$ and $\chi$
are global potentials as in \eq{Ernst},\eq{twistex} and \eq{Komeq},
satisfying~\eq{KFfalloff1}. Let $m$ and $\vec J$ denote the ADM mass
and angular momentum of $M_1$, and let $Q_E$ and $Q_B$ be the global
electric and magnetic charges of $M_1$.
 If~\eqref{Mainineq} holds and $\MUone$ is simply
connected, then
$$
 m \ge \sqrt{\frac{|\vec J|^2}{m^2}+Q_E^2+Q_B^2} \;.
$$
\end{Theorem}

\begin{Remark}
We stress the fact that, in the previous result, no constraints are
assumed. Consequently a clear abuse is created in adopting the
electromagnetic terminology. Nonetheless the parallelism is obvious,
as this result is a ``natural" technical generalization of
Theorem~\ref{TM1}.

For future reference we take the chance to provide formulae for the
``Maxwell" 2-form and the global ``charges" in terms of the {\em
axial} potentials:
 in an orthonormal basis
$\{n,e_i\}$ as before, this time referring to an embedding of $M$ in
a space-time not necessarily satisfying Einstein equations, the
Maxwell 2-form is given by
$$F_{\mu\nu}=
\frac{e^{U}}{\rho}\left(\begin{array}{cccc}
0 & \partial_2\psi & -\partial_1\psi & -\partial_1\chi \\
-\partial_2\psi & 0 &   0 & -\partial_2\chi \\
\;\;\partial_1\psi &  0 &  0  & 0 \\
\;\;\partial_1\chi & \partial_2\chi &  0 & 0
\end{array}\right)\;;
$$
also via equations~\eqref{Ernst} we get
(compare~\cite{Weinstein:1996})
\begin{eqnarray}
 \nonumber
Q_E & = & -\frac{1}{4\pi}\int_{S_{\infty}}*F
=-\frac{-2\pi}{4\pi}\int_{S_{\infty}/U(1)}i_{\eta}*F
\\
& = & \frac{1}{2}\int_{S_{\infty}/U(1)}d\psi
=\frac{\psi_2-\psi_1}{2}\;,
\end{eqnarray}
with a similar computation yielding
$$Q_B=\frac{\chi_1-\chi_2}{2}\;.$$
\end{Remark}

\proof If the mass is infinite there is nothing to prove, otherwise
by \eq{mf2} we need to find a lower bound on
\bel{action} I:=  \int_{\R^3} \Big[ \left(D U\right)^2 +\frac
{e^{4U}} {\rho^4} \left(D \myomega+\chi D\psi - \psi D \chi
\right)^2 +\frac {e^{2U}} {\rho^2} \left((D  \chi)^2+
(D\psi)^2\right)\Big]d^3x
 \;.
\ee

 Let $(\tilde U,\tilde \myomega, \tilde \chi,\tilde \psi)$ be the harmonic map associated with the extreme Kerr-Newman with angular momentum along
the $z$--axis equal to $(\myomega_2-\myomega_1)/8$ and electric
charge $(\psi_2-\psi_1)/2$. We wish to show that the  action
$I:=I(U, \myomega, \chi, \psi)$ is larger than or equal to that of
$(\tilde U, \tilde \myomega, \tilde \chi,\tilde \psi)$, which shall
be denoted by $\tilde I$. As we shall see $I$  differs form an
harmonic map action $H$~\eqref{harmonic map} by a boundary term; the
idea is then to use a result of~\cite{Hildebrandt}, that the action
$H$ is minimized by the solution of the Dirichlet problem which is
expected to be $(\tU,\tilde\myomega,\tilde\chi,\tilde\psi)$;
however, that result does not apply directly because of the
singularity of the equations at the axis $\rho=0$; moreover, we are
working in an unbounded domain. We will overcome such problems by
constructing a  controlled sequence of integrals over compact
domains which avoid the singular set and saturate $\R^3$. Such
strategy was developed in~\cite{CLW}; here we generalize it to the
electro-vacuum setting with considerable simplifications of the
intermediary steps.

So, let $\sigma>0$, $r=\sqrt{\rho^2+z^2}$ and let $f_{\sigma}\in
C^\infty(\R^3)$ be any family of functions satisfying

\begin{itemize}
\item $\partial_{\varphi} f_{\sigma}\equiv0$\;;
\item $0 \le f_{\sigma} \le 1$;
\item $f_{\sigma}=0$ on the set $\{r \le \sigma /2\}\cup\{r\ge
2/\sigma\}$;
 \item
  $f_{\sigma}=1$ on the set $\{ r\ge \sigma\}\cap\{r \le 1/\sigma\}$;
 \item
 $|Df_{\sigma} |\le C/\sigma $ for $\sigma/2 \le r \le
\sigma $\;; and
 \item $|Df_{\sigma}| \le C\sigma $ for $1/\sigma \le r \le
2/\sigma $\;.
\end{itemize}

Let $\theta=U,v,\chi,\psi$ and write

$$\theta_{\jmu }:=f_{\jmu }\theta+(1-f_{\jmu })\,\tilde\theta=f_{\jmu }(\theta-\tilde\theta)+\tilde\theta\;.$$
 We
claim that $I^{\sigma}:=I(U_{\sigma}, \myomega_{\sigma},
\chi_{\sigma},\psi_{\sigma})$ satisfies
\begin{Lemma}
\label{LIeta} $
 \lim_{\,\sigma\to 0} I^{\sigma} = I $.
\end{Lemma}

\proof Indeed, for
\bel{lambdasig}\lambda_{\,\jmu } :=  D v_{\jmu }+\chi_{\jmu } D
\psi_{\jmu }-\psi_{\jmu } D\chi_{\jmu }\;, \ee
we have
\bean \int_{\R^3} \frac{e^{4U_{\jmu}}}{\rho^4} |\lambda_{\,\jmu} |^2
&= & \underbrace{\int_{\{0
 \le \jmu  /2\}} \frac{e^{4 \tilde U}}{\rho^4} |\tilde \lambda |^2}_{I} + \underbrace{\int_{\{\jmu
/2\le r\le \jmu  \}} \frac{e^{4U_{\jmu}}}{\rho^4} |\lambda_{\,\jmu}
|^2}_{II} +
 \\
 \nonumber
 &  & +\underbrace{\int_{\{\jmu \le r \le 1/\jmu  \}} \frac{e^{4U}}{\rho^4} |\lambda |^2}_{III} +
 \underbrace{\int_{\{1/\jmu \le r \le 2/\jmu  \}} \frac{e^{4U_{\jmu}}}{\rho^4} |\lambda_{\,\jmu} |^2}_{IV}%
 \\
 \nonumber
 &  &
 +\underbrace{\int_{\{2/\jmu \le r \}} \frac{e^{4\tilde U}}{\rho^4} |\tilde \lambda |^2}_{V}
 \;.
\eeal{Udecreta}
Since the maps under consideration have finite energy, the integrals
$I$ and $V$ converge to zero, by the dominated convergence theorem.
$III$ converges to the integral over $\R^3$ of $\frac{e^{4
U}}{\rho^4} |\lambda |^2$ by, e.g., the monotone convergence
theorem.
%

We will now show that

\bel{II2} II=\int_{\{\jmu /2\leq r\leq \jmu \}}\frac{e^{4U_{\jmu
}}}{\rho^4}\left( D v_{\jmu }+\chi_{\jmu } D \psi_{\jmu }-\psi_{\jmu
} D\chi_{\jmu }\right)^2\rightarrow_{\jmu \rightarrow 0} 0\;. \ee

The key identity is
\begin{eqnarray}
\label{goodex} \lambda_{\jmu } &=& f_{\jmu }\lambda + (1-f_{\jmu
})\tilde\lambda +  D f_{\jmu }(v-\tilde v)+ D f_{\jmu
}(\tilde\chi\psi-\tilde\psi\chi)
\\
\nonumber && + f_{\jmu }(1-f_{\jmu })\left\{(\psi-\tilde\psi)
D(\chi-\tilde\chi)-(\chi-\tilde\chi) D(\psi-\tilde\psi)\right\}\;,
\end{eqnarray}
which will allow to establish~\eqref{II2} by a step-by-step
estimation of the integrals obtained by replacing $\lambda_{\,\jmu
}$ by each of its five summands. We will exemplify this by dealing
with the most delicate case.

\newcommand{\jalpha}{\beta}
\newcommand{\les}{\lesssim}

The existence of multiple ends manifests itself in the asymptotic
behavior
\bel{asympPunc2}
 U  = 2 \log r+
O(1) \;,\,\,r\rightarrow 0\;,
\ee
established in~\cite[Theorem~2.6]{ChUone}. Then, using the decay
rates of the extreme Kerr-Newman map, compiled in
Table~\ref{table:EKN} of the appendix, we get~\footnote{We will
write $f\les g$ if and only if $f=O(g)$.}
\begin{eqnarray}
\label{Uzero} \nonumber e^{4U_{\jmu }}&=&e^{4f_{\jmu
}U}e^{4(1-f_{\jmu })\tilde U}
\\
\nonumber &\les& r^{8f_{\jmu }}r^{4(1-f_{\jmu })}
\\
&=& r^{4(f_{\jmu }+1)}\leq r^4\;,\,\,r\rightarrow 0.
\end{eqnarray}
Recall that near $r=0$  the coordinates $(\rho,z)$ can be obtained
from the usual cylindrical coordinates in the other asymptotically
flat region, which we denote by $(\hat \rho, \hat z)$, by an
inversion $(\hat\rho,\hat z)=(\frac{\rho}{r^2},\frac{z}{r^2})$,
compare~\cite[Theorem~2.9, p.~2580]{ChUone}. This leads to estimates
for small $r$, equivalently for large $\hat r$, such as

\bel{lambda origin} |\lambda|_{\delta}=
\frac{1}{r^2}|\lambda|_{\hat\delta}\les \frac{1}{r^2}\hat\rho^2\hat
r^{-\jalpha } \les \frac{1}{r^2}\frac{\rho^2}{r^4}r^{\jalpha
}=\rho^2 r^{\jalpha -6}\;,\,\,r\rightarrow 0\;,\ee
where $\hat\delta=d\hat{\rho}^2+d\hat z^2+\hat{\rho}^2d\varphi^2$.

The same procedure yields, from the estimates in
Table~\ref{table:EKN},
\bel{dtheta origin} |D\chi|_{\delta},|D\psi|_{\delta}=\rho
O(r^{\gamma-3})\;,\,\,r\rightarrow 0\;, \ee
and we see that, for small $r$,
\bel{dv origin} |D v|_{\delta} \leq |\lambda|_{\delta}+|\chi
D\psi-\psi D\chi|_{\delta} \les \rho^2 r^{\beta-6} + \rho
O(r^{2\gamma-4})\;.
\ee
From this and the known asymptotic behaviour of extreme Kerr-Newman
one obtains, when  $\beta\ge 2\gamma+1$,%
\footnote{For $\beta< 2\gamma+1$ the dominating behaviour
in~\eqref{dv origin} is governed by $\lambda$, which leads to
$v-\tilde v=O(r^{\beta-3})$ and the necessity to impose $\beta>5/2$,
as in vacuum~\cite[p.~2602]{CLW}.}
\bel{v-v origin} v-\tilde v=O(r^{2\gamma-2})\;,\ee
and we are now able
 to see that the contribution of the term $ D f_{\mu}(v-\tilde v)$ in the region $\rho\geq z$, where $r$ is comparable to $\rho$, is estimated by
$$
 \int_{\sigma/2}^{\sigma}\frac{r^4}{\rho^4r^2}\left(r^{2\gamma-2}\right)^2r^2dr
 \les \sigma^{4\gamma-3}\to_{\sigma\to 0} 0 \ \text{provided that} \ \gamma>3/4
\;.
$$
This explains our ranges of $\beta$ and $\gamma$ in~\eqref{falloff1}
and~\eqref{KFfalloff1}.

Since $v$ and $\tilde v$ have the same axis data, Taylor expanding
on $\rho$ along the axis yields
\bel{Taylor v-v} (v-\tilde v)(\rho,z)=\underbrace{(v-\tilde
v)(0,z)}_{=0}+\partial_{\rho}(v-\tilde v)(c(\rho),z)\rho\;,\,
|c(\rho)|\leq |\rho|\,. \ee
Also, again for $\beta\ge 2\gamma+1$,
$$
\partial_{\rho}v=\rho O(r^{2\gamma-4})\;,$$
with the same estimate holding for the $\rho$-derivative of the
difference. Then, in $\{\rho\leq z\}$,
\bel{v-v origin2} v-\tilde v= \rho^2
O(r^{2\gamma-4})\;,\,\,r\rightarrow 0\;.
\ee
We see that, in this region, the integral under consideration is
estimated by
$$\int_{\{\theta:\rho<z\}}\int_{\sigma/2}^{\sigma}\frac{r^4}{\rho^4r^2}\left(\rho^2r^{2\gamma-4}\right)^2r^2\sin\theta\,drd\theta
\les \sigma^{4\gamma-3}\rightarrow_{\sigma\rightarrow 0} 0\;,$$
and~\eqref{II2} follows.

The remaining terms in $I^{\sigma}$ can be controlled in a similar,
although considerably more direct and simpler, fashion. For
instance, when controlling the  $|DU_{\sigma}|^2$ term one of the
steps requires to estimate the integral
\beaa
  \int_{\{\sigma/2 \le r \le \sigma  \}}  |D U_\sigma|^2
 &= & \int_{\{\sigma/2 \le r \le \sigma  \}} { |(U   - \tU) D f_{\sigma} + f_{\sigma} D
U  +(1-f_{\sigma}) D \tU|^2}
  \\
 & \les & \int_{\{\sigma/2 \le r \le \sigma  \}}
 \left( {(U   - \tU)^2 r^{-2}} + |
 D U  |^2+ |  D \tU|^2\right)
 \;,
\eeaa
where in fact the second and third term go to zero by the Lebesgue
dominated convergence theorem while the vanishing of the first
follows by direct estimation using~\eqref{asympPunc2} and the decay
rates presented in Table~\ref{table:EKN}.
 \qed

We now show that:

\begin{Lemma}
\label{IetaItilde} $I^{\sigma} \ge \tilde I$ for all $\sigma$ small
enough.
\end{Lemma}

\proof This time consider, for $0<\epsilon<1$,
\newcommand{\hf}{\hat f_{\epsilon}}
$$
\hf  = \left\{
                 \begin{array}{ll}
                   0, & \hbox{$\rho \le {\epsilon}$}\,; \\
                   \frac{\log\frac{\rho}{\epsilon}}{\log \frac{\sqrt\epsilon}{\epsilon}}, & \hbox{$\epsilon \le  \rho \le \sqrt\epsilon$\,;} \\
                   1, & \hbox{$\rho \ge \sqrt\epsilon$\,.}
                 \end{array}
               \right.
$$

Set, for $\theta=U,v,\chi,\psi$,
\beaa \theta_{\jmu ,\epsilon} = \hf \theta_\jmu  + (1-\hf )
\tilde\theta\;, &
 \eeaa
and let $I^{\jmu ,\epsilon} $ denote the action of $(U_{\jmu
,\epsilon} , v_{\jmu ,\epsilon}, \chi_{\jmu ,\epsilon}, \psi_{\jmu
,\epsilon} )$ and
$$\lambda_{\jmu ,\epsilon}=D v_{\jmu ,\epsilon}+\chi_{\jmu ,\epsilon}D\psi_{\jmu ,\epsilon}-\psi_{\jmu ,\epsilon}D\chi_{\jmu ,\epsilon}\;.$$
We claim that
\bel{Itend} \int_{
{\{\rho \le \sqrt\epsilon\}}} \Big[ \left(D U_{\jmu
,\epsilon}\right)^2 +\frac {e^{4 U_{\jmu ,\epsilon}}} {\rho^4}
\left(\lambda_{\jmu ,\epsilon}\right)^2+ \frac {e^{2 U_{\jmu
,\epsilon}}} {\rho^2} \left((D\chi_{\jmu ,\epsilon})^2+(D\psi_{\jmu
,\epsilon})^2 \right)\Big]d^3x\to_{\epsilon\to 0} 0 \;.
 \ee
Equivalently,
\bel{Itend2} I^{\jmu ,\epsilon} \to_{\epsilon\to 0} I^{\jmu} \;.
 \ee
In order to see this, note that the integral over the set $\{0\le
\rho \le \epsilon\}$, where $\theta_{\jmu ,\epsilon}=\tilde \theta$,
approaches zero as $\epsilon$ tends to zero by the Lebesgue
dominated convergence theorem; the same  happens away from the set
$\{\jmu /2<r<2/\jmu \}$. So it remains to consider the integral over
$$
\mcW_{\jmu ,\epsilon}:=
\{\epsilon\le \rho \le \sqrt\epsilon\}\cap \{\jmu /2<r<2/\jmu\}
 \;.
$$

 The computations leading
 to~\eqref{goodex} now give
\begin{eqnarray}
\label{goodex2} \lambda_{\jmu ,\epsilon} &=& \hf\lambda_{\jmu } +
(1-\hf)\tilde\lambda +  D \hf(v_{\jmu }-\tilde v)+ D
\hf(\tilde\chi\psi_{\jmu }-\tilde\psi\chi_{\jmu })
\\
\nonumber && + \hf(1-\hf)\left\{(\psi_{\jmu }-\tilde\psi)
D(\chi_{\jmu }-\tilde\chi)-(\chi_{\jmu }-\tilde\chi) D(\psi_{\jmu
}-\tilde\psi)\right\}\;.
\end{eqnarray}

Since $I^{\jmu}\rightarrow I$, we see that $I^{\jmu}$ must be
finite, at least for all small enough $\jmu $. Fix such a $\jmu >0$.
As before the first two terms in the right-hand side
of~\eqref{goodex2} constitute no problem. To control the others note
that, for all $\epsilon$ such that $\sqrt{\epsilon}<\jmu /2$, we
have, in the $(\rho,z,\varphi)$ coordinates,
\bel{W} \mcW_{\jmu ,\epsilon}\subseteq
[\epsilon,\sqrt{\epsilon}\,]\times\left([z_0(\jmu ),z_1(\jmu
)]\cup[z_2(\jmu ),z_3(\jmu )]\right)\times[0,2\pi]\;, \ee
for a good choice of $z_i$'s satisfying $z_i(\jmu )\neq 0$; e.g.,
the $z$--coordinate value, in increasing order, of the points in the
intersection of $\rho=\epsilon$ with both $r=\sigma/2$ and
$r=2/\sigma$. We then see that
\begin{eqnarray*}
\int_{\mcW_{\jmu ,\epsilon}} \frac {e^{4 U_{\jmu ,\epsilon}}}
{\rho^4} \left(D \hf(v_{\jmu}-\tilde v)\right)^2 d^3x &\leq&
2\pi\sum_{i=0,1}\int_{z_{2i}(\jmu )}^{z_{2i+1}(\jmu
)}\int_{\epsilon}^{\sqrt{\epsilon}} \frac {e^{4 U_{\jmu ,\epsilon}}}
{\rho^4} \left(D \hf\right)^2\left(v_{\jmu}-\tilde v\right)^2\rho\,
d\rho\, dz
\\
&\leq& 2\pi\sum_{i=0,1}\int_{z_{2i}(\jmu )}^{z_{2i+1}(\jmu
)}\int_{\epsilon}^{\sqrt{\epsilon}} \frac {C(\jmu )}
{\rho^3}\frac{1}{\rho^2(\log\epsilon)^2} \left(v_{\jmu}-\tilde
v\right)^2 d\rho\, dz
\end{eqnarray*}

Since $z_i(\jmu )\neq 0$ we see that $v_{\jmu }$ and $\tilde v$ are
smooth on the set $\{\rho\leq \sqrt{\epsilon}\;,\,\,
z\in\cup_i[z_{2i},z_{2i+1}]\}$. Then, Taylor expanding on $\rho$
along the axis, while noting that $v_{\jmu }$ and $\tilde v$ have
the same axis data and that $f_{\sigma}$ is, by construction,
axisymmetric, yields (compare~\eqref{Taylor v-v})
\bel{another Taylor} v_{\jmu }-\tilde v=O(\rho^2)
\;,\,\,\rho\rightarrow 0\;,\text{ in } \{\rho\leq
\sqrt{\epsilon}\;,\,\, z\in\cup_i[z_{2i},z_{2i+1}]\}\;,\ee
hence
\begin{eqnarray*}
2\pi\sum_{i=0,1}\int_{z_{2i}(\jmu )}^{z_{2i+1}(\jmu
)}\int_{\epsilon}^{\sqrt{\epsilon}} \frac {C(\jmu )}
{\rho^3}\frac{1}{\rho^2(\log\epsilon)^2} \left(v-\tilde v\right)^2
d\rho\, dz &\les& \frac{C(\jmu
)}{(\log\epsilon)^2}\int_{\epsilon}^{\sqrt{\epsilon}} \frac {1}
{\rho^5} \rho^4\, d\rho
\\
&\les& \frac{C(\jmu )}{(\log\epsilon)^2} \log\epsilon
\rightarrow_{\epsilon\rightarrow 0} 0\;.
\end{eqnarray*}

The remaining terms are controlled in an analogous way, with the
$(DU_{\jmu ,\epsilon})^2$ term behaving exactly as in
vacuum~\cite{CLW}. This  ends the proof of \eq{Itend}.

\bigskip

Using the rescaling  $U=u+\ln\rho$ we have
\bel{action harmonic}
I_{\Omega}(U,v,\chi,\psi)=H_{\Omega}(u,v,\chi,\psi)+B_{\Omega}(U)\;,
\ee
where
\bel{harmonic map} H_{\Omega}= \int_{\Omega} \Big[ \left(D
u\right)^2 +{e^{4u}} \left(D \myomega+\chi D\psi - \psi D \chi
\right)^2 + {e^{2u}}  \left((D \chi)^2+ (D\psi)^2\right)\Big]d^3x\;,
\ee
is the energy of the harmonic map
\bel{harmonic map0}
 \Phi=(u,v,\chi,\psi):\R^3\setminus \mcA
 \longrightarrow \mathbb{H}^2_{\mathbb{C}}\;,
\ee
which differs from $I$ by the boundary term
\bel{boundary term} B_{\Omega}(U)=\int_{\partial
\Omega}\frac{\partial\ln\rho}{\partial N}(2U-\ln\rho)dS\;, \ee
where $N$ is the outward pointing unit normal to $\partial \Omega$.
Consequently for both $I$ and $H$  the associated variational
equations are the harmonic map equations, with target space the
two-dimensional complex hyperbolic space. Hence the target manifold
satisfies the convexity conditions of~\cite{Hildebrandt} (see Remark
(i), p.~5 there). For compact $\Omega$ away from the axis we can
thus conclude from~\cite{Hildebrandt} that action minimisers of
$H_{\Omega}$ with Dirichlet boundary conditions exist, are smooth,
and satisfy the variational equations. It is also well known
(see~\cite{CLW} and references therein) that solutions of the
Dirichlet boundary value problem are unique when the target manifold
has negative sectional curvature, which is the case here. All this
implies that $(\tU,\tilde\myomega,\tilde\chi,\tilde\psi)$, with its
own boundary data, minimizes the action integral $H$~, and
consequently of $I$, over the sets
\newcommand{\Cse}{\mycal{C}_{\sigma,\epsilon}}
$$
 \Cse:={\{\rho \ge \epsilon\}}\cap \{\sigma/2\le r\le 2/\sigma\}
 \;.
$$
In particular, since  the maps $(\theta_{\sigma,\epsilon})$ and
$(\tilde\theta)$ coincide on $\partial\Cse$
we conclude that
$$I_{\Cse}(U_{\sigma,\epsilon} , \myomega_{\eta,\epsilon}, \chi_{\eta,\epsilon}, \psi_{\eta,\epsilon} )
\ge I_{\Cse}(\tilde U,\tilde v, \tilde \chi,\tilde\psi)\;. $$
In fact the maps under consideration coincide on the closure of the
complement of $\Cse$ and therefore
\bel{main} I^{\sigma,\epsilon}\ge\tilde I\;. \ee

Recalling~\eqref{Itend2} we obtain
\bel{last} I^{\sigma} = \lim_{\epsilon\rightarrow
0}I^{\sigma,\epsilon}
 \geq  \lim_{\epsilon\rightarrow 0}\tilde I=\tilde I\;.
 \ee
 \qed

\medskip

Returning to the proof of Theorem~\ref{TM1}, Lemmata~\ref{LIeta}
and~\ref{IetaItilde} yield
$$I=\lim_{\,\sigma\to 0} I^{\sigma}\geq \tilde I\;,$$
and the result is a consequence of Remark~\ref{m=I} followed by a
duality rotation.
 \qed

\section{Concluding remarks}
 \label{sCrDain}

The study of Dain inequalities is still in an early stage with
important questions still needing to be settled even for pure
vacuum; also, some impressive generalization can be easily
formulated and justified by the heuristic argument presented in the
introduction. We finish this chapter by addressing some of this
issues:

\begin{enumerate}
\item {\bf Extreme Kerr-Newman as a minimum of the mass functional.}
Our class of data does not include extreme Kerr-Newman  and
consequently eliminates a priori the possibility of establishing it
as the unique minimum for the mass functional. This difficulty is
not present in Dain's original work where a class of Brill data is
considered; however, this is done at the cost of a considerably
longer list of (stronger) technical assumptions, some of which are
derivable properties of asymptotic flatness and the existence of
multiple ends as was first observed and established by Chru\'sciel
in~\cite{ChUone}. To obtain the desired result within the spirit of
the program initiated by Chru\'sciel one could start by generalizing
the results in~\cite{ChUone} for data allowing for both
asymptotically flat and asymptotically cylindrical ends, and then
try to adapt the arguments presented here.

From what as been said, we expected inequality~\eqref{21VI.2} to be
strict within the class of data considered in this work.

\item {\bf Multiple asymptotically flat ends.} Even for vacuum the
question of multiple ends requires further work. Although a Dain
inequality was already established in~\cite{CLW} it depends on a
function of the angular momenta for which an explicit expression
remains unknown for all $N>2$, where $N$ is the number of
asymptotically flat ends. This is clearly an unsatisfactory
situation since  the Penrose-like argument, presented at the
beginning of this paper, provides evidence that the unknown function
should  simply be the square root of the total angular momentum. In
fact,  for the two body problem, $N=3$, such expectation as been
recently supported by numerical evidences~\cite{Dain:2009}.

One also expects the ideas in~\cite{CLW} to generalize to
electro-vacuum by using the methods developed here, but in this case
it seems hard to speculate what the exact expression for the lower
bound function, this time of both angular momenta and Maxwell
charges, should be. This problem is related to the fact that the
Majumdar-Papapetrou metrics provide the existence of regular and
extreme multiple black hole solutions; analogous difficulties have
been found for the Penrose inequality~\cite{WeinsteinYamada}.

\item {\bf Asymptotically electro-vacuum data.} The heuristic
argument leading to the Dain inequality presented here works for
other data, involving far more general matter models:  axisymmetric
asymptotically electro-vacuum initial data whose domain of outer
communications becomes electro-vacuum asymptotically with time.
Establishing a Dain inequality in such a generality would be quite
impressive but, at this moment, such goal seems unreachable.

\end{enumerate}
\bigskip

{\sc Acknowledgements:} The author is grateful to Piotr Chru\'sciel
for making him a part of this challenging project.

\appendix
\section{Decay rates for Extreme Kerr-Newman}
\label{EKN}

\begin{table}[ht]
\centering 
\begin{tabular}{c c} 
\hline\hline 
$r\rightarrow 0$ & $r\rightarrow +\infty$ \\
\hline  
\\
$\tilde U=\log(r)+O(1)$   \;\;\;\;    &   \;\;\;\;$\tilde
U=-\frac{m}{r}+O(r^{-2})$
\\ [1ex]
$\tilde\chi=\rho^2O(r^{-2})=O(1)$   \;\;\;\;    &   \;\;\;\;
$\tilde\chi=\rho^2O(r^{-3})=O(r^{-1})$
\\ [1ex]
$\tilde\psi=O(1)$                           \;\;\;\;    &
\;\;\;\;$\tilde\psi=\rho O(r^{-2})=O(r^{-1})$
\\ [1ex]
$\partial_{\rho}\tilde\chi=\rho O(r^{-2})$  \;\;\;\;    &   \;\;\;\;  $\partial_{\rho}\tilde\chi=\rho O(r^{-3})$
\\ [1ex]
$\partial_{\rho}\tilde\psi=\rho O(r^{-2})$  \;\;\;\; &  \;\;\;\;
$\partial_{\rho}\tilde\psi=\rho
 O(r^{-2})$
\\ [1ex]
$|D\tilde\chi|_{\delta}=\rho O(r^{-2})$  \;\;\;\; & \;\;\;\;
$|D\tilde\chi|_{\delta}=\rho O(r^{-3})$
\\  [1ex]
$|D\tilde\psi|_{\delta}=O(r^{-1})$  \;\;\;\; &  \;\;\;\;
$|D\tilde\psi|_{\delta}=O(r^{-1})$
\\  [1ex]
$\tilde v=O(1)$ \;\;\;\; & \;\;\;\; $\tilde v=O(1)$
\\  [1ex]
 $\partial_{\rho}\tilde v=\rho
O(r^{-2})$ \;\;\;\; & \;\;\;\; $ \partial_{\rho}\tilde v=\rho
O(r^{-2})$
\\ [1ex]
 \hline
\end{tabular}
\caption{Decay rates for Extreme Kerr-Newman} 
\label{table:EKN} 
\end{table}

\def\cprime{$'$} \def\cprime{$'$} \def\cprime{$'$} \def\cprime{$'$}
\providecommand{\bysame}{\leavevmode\hbox
to3em{\hrulefill}\thinspace}
\providecommand{\MR}{\relax\ifhmode\unskip\space\fi MR }
\providecommand{\MRhref}[2]{%
  \href{http://www.ams.org/mathscinet-getitem?mr=#1}{#2}
} \providecommand{\href}[2]{#2}

\end{document}